\documentclass[runningheads]{llncs}
\usepackage{graphicx}
\usepackage{todonotes}
\usepackage[colorlinks=true,
			urlcolor=blue,
			linkcolor=black,
			citecolor=black]{hyperref}  %
\usepackage{amsmath}  %
\usepackage{listings}
\usepackage{xcolor}
\usepackage{wrapfig}
\usepackage{tikz}

\definecolor{codegreen}{rgb}{0,0.6,0}
\definecolor{codegray}{rgb}{0.5,0.5,0.5}
\definecolor{codepurple}{rgb}{0.58,0,0.82}
\definecolor{backcolour}{rgb}{0.97,0.97,0.97}

\lstdefinestyle{mystyle}{
    backgroundcolor=\color{backcolour},   
    commentstyle=\color{codegreen},
    keywordstyle=\color{magenta},
    numberstyle=\tiny\color{codegray},
    stringstyle=\color{codepurple},
    basicstyle=\ttfamily\footnotesize,
    breakatwhitespace=false,         
    breaklines=true,                 
    captionpos=b,                    
    keepspaces=true,                 
    numbers=left,                    
    numbersep=5pt,                  
    showspaces=false,                
    showstringspaces=false,
    showtabs=false,                  
    tabsize=2
}

\begin{document}

\title{Cucumber: Renewable-Aware Admission Control for Delay-Tolerant Cloud and Edge Workloads}
\titlerunning{Cucumber: Renewable-Aware Admission~Control}

\author{
Philipp Wiesner\inst{1} \and %
Dominik Scheinert\inst{1} \and %
Thorsten Wittkopp\inst{1} \and %
\\
Lauritz Thamsen\inst{2} \and %
Odej Kao\inst{1} %
}

\authorrunning{P. Wiesner et al.}

\institute{
Technische Universität Berlin, Berlin, Germany\\
\email{\{wiesner, dominik.scheinert, t.wittkopp, odej.kao\}@tu-berlin.de}
 \and
University of Glasgow, Glasgow, United Kingdom\\
\email{lauritz.thamsen@glasgow.ac.uk}
}

\maketitle              %

\begin{tikzpicture}[remember picture,overlay]
    \node[anchor=south,yshift=40pt] at (current page.south) {\parbox{\dimexpr\textwidth-\fboxsep-\fboxrule\relax}{
        \scriptsize 28th International European Conference on Parallel and Distributed Computing (Euro-Par 2022) \\
        \textcopyright 2022 Springer | LNCS, volume 13440 | DOI: \href{https://link.springer.com/chapter/10.1007/978-3-031-12597-3_14}{10.1007/978-3-031-12597-3\_14}}
    };
\end{tikzpicture}

\begin{abstract}

The growing electricity demand of cloud and edge computing increases operational costs and will soon have a considerable impact on the environment.
A possible countermeasure is equipping IT infrastructure directly with on-site renewable energy sources.
Yet, particularly smaller data centers may not be able to use all generated power directly at all times,
while feeding it into the public grid or energy storage is often not an option.
To maximize the usage of renewable excess energy, we propose Cucumber, an admission control policy that accepts delay-tolerant workloads only if they can be computed within their deadlines without the use of grid energy.
Using probabilistic forecasting of computational load, energy consumption, and energy production, Cucumber can be configured towards more optimistic or conservative admission.
We evaluate our approach on two scenarios using real solar production forecasts for Berlin, Mexico City, and Cape Town in a simulation environment. 
For scenarios where excess energy was actually available, our results show that Cucumber's default configuration achieves acceptance rates close to the optimal case and causes 97.0\,\% of accepted workloads to be powered using excess energy, while more conservative admission results in 18.5\,\% reduced acceptance at almost zero grid power usage.

\end{abstract}

\section{Introduction}
\label{sec:intro}

As the demand for computing continues to grow year by year, so are operating expenses and the associated carbon emissions caused by consuming energy from the public power grid~\cite{Freitag_ICTClimateImpact_2021}.
So far, negative effects could partially be mitigated through advances in hardware efficiency, cooling, and the continuous shift of cloud computing towards highly energy-optimized hyperscale data centers, which already host about 50\,\% of all compute instances~\cite{Masanet_RecalibratingGlobalDCEnergyEstimates_2020}.
Still, data centers already account for more than 1\,\% of global energy consumption and this number is expected to rise further~\cite{Masanet_RecalibratingGlobalDCEnergyEstimates_2020} -- especially when considering the additional demand of novel domains like the internet of things (IoT), edge and fog computing~\cite{Wiesner_LEAF_2021}.

\begin{figure}
    \centering
       \includegraphics[width=0.9\textwidth]{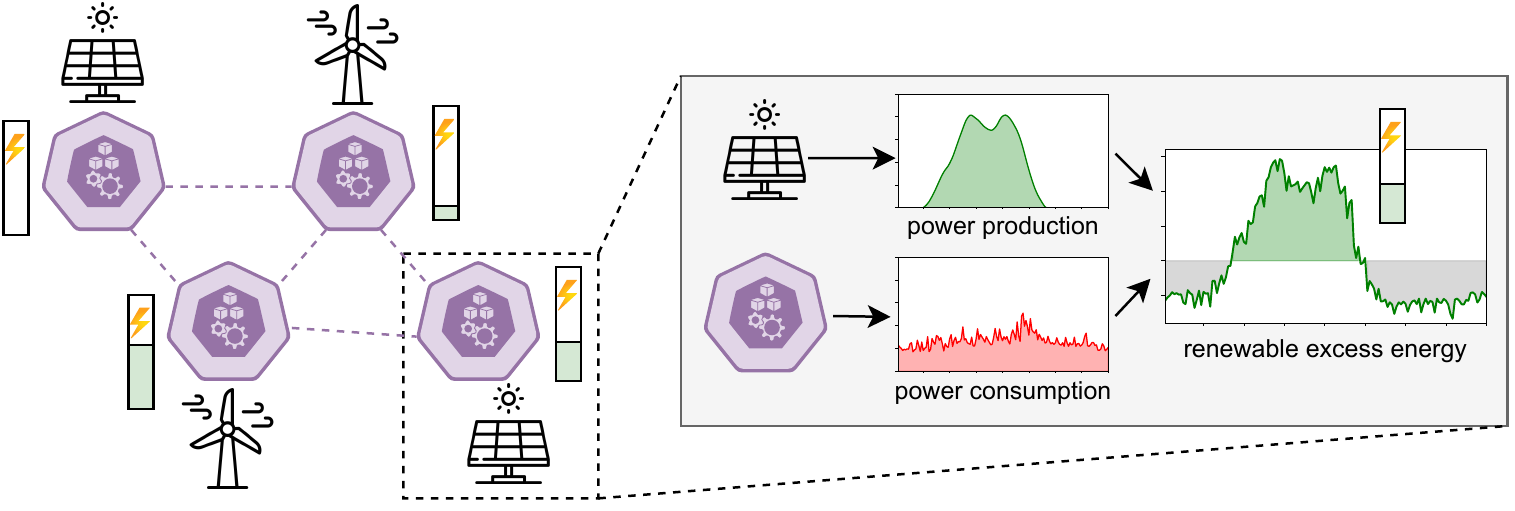}
    \caption{Problem setting: Renewable excess energy can ocurr at compute nodes when local demand does temporarily not cover all produced energy. %
    }
    \label{fig:problem_overview}
\end{figure}

To reduce its carbon footprint, the IT industry is pushing to integrate more and more low-carbon energy sources into data centers~\cite{acun2022holistic}, not least because carbon pricing mechanisms, such as emission trading systems or carbon taxes, are starting to be implemented around the globe \cite{WorldBank_CarbonPricing_2020}.
For example, Google plans to operate their data centers solely on carbon-free energy by 2030 \cite{Google_CarbonFreeBy2030_2020}.
One approach towards more sustainable and cost-effective computing systems in cloud as well as edge environments is directly equipping IT infrastructure with on-site renewable energy sources like solar or wind~\cite{SDIA_OnsitePowerDataCenters_2021,LiYDPTKZ18}. %
However, especially smaller compute nodes, such as on-premise installations or edge data centers, are not always able to consume all generated power directly, as depicted in \autoref{fig:problem_overview}.

\vspace{0.2cm}

Energy storage can mitigate this problem to some extent, but is expensive, therefore often not available in sufficient capacity, and may be reserved to ensure operation during power outages.
Moreover, storing energy involves power conversion loss, and frequent charging cycles accelerate battery aging~\cite{Liu_BatteryAging_2017}.
On the other hand, feeding excess energy back to the power grid is often unattractive in practice due to statutory regulations and low compensation.
Microgrids address this by directly integrating renewables and energy storage to locally balance excess energy~\cite{Hirsch_Microgrids_2018}.
Such systems can greatly benefit from participants who are flexible and able to adapt their energy consumption to the expected supply.

To make better use of renewable excess energy (REE) occurring close to compute nodes, delay-tolerant workloads originating locally or within the surrounding distributed system should be computed on free computational capacity.
Delay-tolerant workloads are common in cloud environments, ranging from machine learning jobs, certain Function-as-a-Service (FaaS) executions, nightly backups and CI/CD runs, and other periodic jobs like generating daily reports~\cite{Wiesner_LetsWaitAwhile_2021}.
However, they may also occur in otherwise time-critical edge computing environments, such as cache and index updates as well as federated and/or iterative machine learning trainings on locally available data at edge nodes.

We propose Cucumber, an admission control policy for delay-tolerant workloads in resource-constrained compute nodes that have access to renewable energy sources but no access to energy storage.
We assume that this infrastructure usually runs high-priority, time-critical workloads with quality of service (QoS) constraints, like user-facing services, but is not always fully utilized.
Cucumber admits delay-tolerant workloads to the local system only if they can be computed within their deadlines on free capacity and without the use of grid energy.
This leads to increased use of renewable energy sources, hence reducing associated carbon emissions and electricity costs, and contributes to stabilizing the power grid. 
We furthermore expect Cucumber to be an integral building block of decentralized systems that exploit the varying spatio-temporal  availability of renewable energy.
Towards this, we make the following contributions:

\begin{itemize}
	\item we define a method for forecasting free computational capacity that can be powered using REE only. The prediction can be tuned towards conservative or optimistic results using probabilistic forecasts of load, energy consumption and energy production
	\item based on these forecasts, we propose an admission control policy that decides whether incoming delay-tolerant workloads with known size and deadline can be scheduled on free capacity using REE only
	\item we evaluate our approach on two scenarios using real solar production forecasts for Berlin, Mexico City, and Cape Town in a simulation environment
	\item we make all datasets and code used for this experimental evaluation publicly available for future research to build on our results\footnote{Github: \url{https://github.com/dos-group/cucumber}}
\end{itemize}

The remainder of this paper is structured as follows:
\autoref{sec:related_work} reviews related work.
\autoref{sec:apporach} proposes the admission control policy and explains how we generate forecasts on free computational capacity that can be powered by REE.
\autoref{sec:evaluation} evaluates our approach.
\autoref{sec:conclusion} concludes the paper.

\section{Related Work}
\label{sec:related_work}

\paragraph{Carbon-Aware and Renewable-Aware Computing.}

Incorporating the availability of renewable or low-carbon power into scheduling decisions 
has been increasingly researched over the last decade.
However, many works in this context focus on load migration in geo-distributed settings or optimize for low carbon signals in the public power grid.
For example, Google employs a suite of analytics pipelines %
to defer delay-tolerant workloads if power from the public grid is associated with high carbon intensity~\cite{Radovanovic_Google_2021}.
While their work is targeted at large-scale data centers, Cucumber is meant to be deployed in resource-constrained environments with direct access to renewable energy sources.
Toosi et al.~\cite{ToosiQAB17noaprioriknowledge} proposed a load balancer for web applications that increases on-site renewable energy usage at data centers.
However, other than Cucumber, their approach is reactive and does not make use of forecasting for better decisions.
GreenSlot~\cite{Goiri_MatchingRenewableEnergyGreenDatacenters_2015} is a batch job scheduler for data centers with on-site renewable energy sources using hourly predictions of solar energy and optimizes for a low price if grid power usage is unavoidable.
Cucumber, on the other hand, aims at using REE only and tries to avoid using any grid power.
Aksanli et al.~\cite{Aksanli_GreenEnergyPredictionScheduleBatchServiceJobs_2011} proposed a scheduler using short-term predictions of wind and solar power to reduce grid power usage and the number of canceled jobs.
In contrast, Cucumber rejects workloads in danger of violating their deadlines upfront so they can be scheduled elsewhere.
The Zero-Carbon Cloud~\cite{chien2019zero} is the only project of which we are aware that aims at exploiting REE by placing data centers close to renewable energy sources.
Our approach complements these efforts and opens up a way to distribute workloads in a decentralized manner across their proposed infrastructure by making local decisions about whether or not to accept a job.

\vspace{-0.1cm}
\paragraph{Admission Control}

is a validation process in communication systems to check if sufficient resources are available to, for example, establish a network connection or process a request.
Other than most publications on admission control that operates on a network packet level, we consider workloads that can be several minutes or even hours long.
Because of this, most related work is in the context of web-based applications or cloud computing where certain requests are prioritized to improve quality of service (QoS) or maximize revenue~\cite{ChenMC01,YuanBTL16}.
An admission control policy in green computing was proposed by~\cite{Hazemi13}, where a PID controller used in industrial control applications is extended by a hybrid green policy, to reduce grid power usage.
Eco-IDC~\cite{LuoRL14} targets energy-aware admission control on a data center level by exploiting electricity price changes while dropping excessive workload if required. %
Other than these approaches, Cucumber %
optimizes for utilizing locally available REE while prioritizing time-critical workloads.
Furthermore, our approach utilizes probabilistic forecasting methods to be configurable towards more optimistic or conservative admission.

\section{Admission Control}\label{sec:apporach}

\begin{figure}
    \centering
    \includegraphics[width=\textwidth]{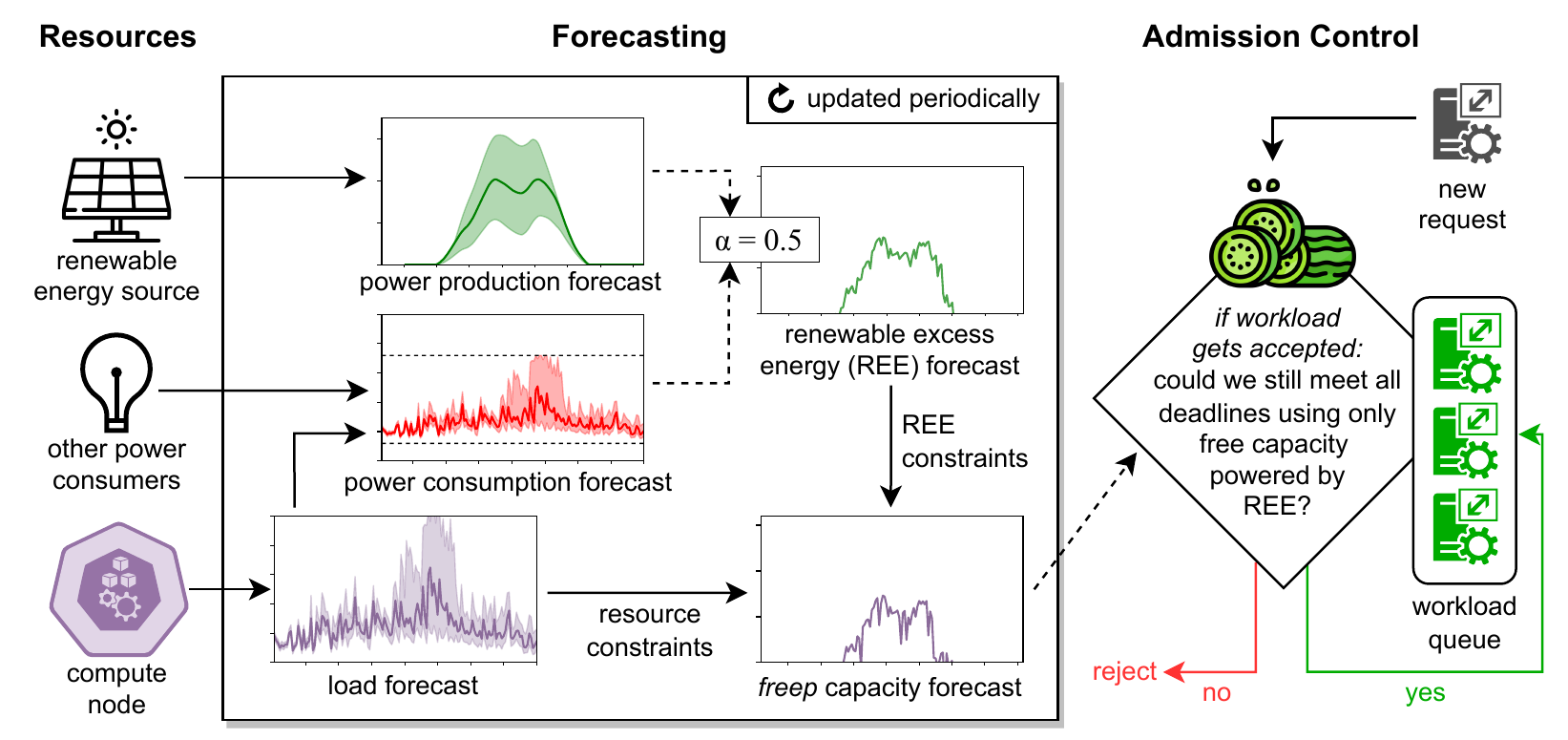}
    \caption{Cucumber periodically forecasts computational load, power consumption, and power production to compute the \emph{freep} capacity forecast. It determines how much computational capacity will be available in the future, that can be powered using REE only. Based on this forecast and the amount, size, and deadlines of already queued workloads, Cucumber accepts or rejects new workload requests.}
    \label{fig:approach_overview}
\end{figure}

Cucumber accepts delay-tolerant workloads based on forecasts of load, power consumption, and power production.
A high-level overview and outline of the approach are presented in \autoref{fig:approach_overview}.
This section describes all steps in detail. %

\subsection{Forecasting Load, Power Consumption, and Power Production}

Cucumber uses probabilistic multistep-ahead forecasts to predict time series of probability distributions, which inherent the uncertainty for each observation, to later infer the available REE at different confidence intervals.
If no probabilistic forecasts are available, Cucumber can still be operated in its default configuration based on the expected/median forecast.

\paragraph{Forecasting Computational Load.}

Load prediction is a widely researched field covering forecasts related to application metrics, such as the number of messages in a stream processing system~\cite{gontarska2021evaluation}, as well as the utilization of (virtualized) hardware resources like CPU, GPU or RAM.
Although load prediction systems are usually formulated as time series forecasting problems based on historical data, they can also take information from other contexts into account.
For example, in edge computing use cases like traffic monitoring, additional information on weather, holidays, events, etc. can improve the forecast quality. %
Whatever type of forecast is most suitable in a concrete use case, Cucumber uses it to identify future time windows with free capacity.
Furthermore, these load predictions are used as a factor in the power consumption forecast.
In the following, we denote the load of a node as $U$ and any load forecasts as $U_\text{pred}$.

\paragraph{Forecasting Power Consumption.}
The power demand of IT infrastructures can be influenced by many factors like CPU or GPU usage, memory, I/O, and storage access, temperature, etc. 
While perfect modeling without precise knowledge of workload and infrastructure characteristics is not possible~\cite{Koller_WattApp_2010}, it has been shown that power usage can often be modeled with sufficient accuracy based only on the node's utilization~\cite{Barroso_EnergyProportionalComputing_2007} -- which usually refers to its CPU usage.
In fact, power modeling based on CPU usage only is being used in production at modern hyper-scale data centers~\cite{Radovanovic_PowerModeling_2021}.
For simplicity, here we assume a simple linear power model to convert from a certain load $U$ to the nodes power usage $P$:
\begin{equation}
P = P_\text{static} + U \cdot (P_\text{max} - P_\text{static})
\label{eq:power}
\end{equation}
where $P_{static}$ is the amount of power a node consumes in idle state and $P_{max}$ is the amount of power the node consumes under full load.
Besides energy used for computing, the power forecast should also take the expected demand from other co-located consumers into account that are powered by the renewable energy source, like cooling or lighting, to correctly derive the actually available REE.

\paragraph{Forecasting Power Production.}

Since information on future power production is useful in many domains ranging from high-level application design to low-level grid control, the prediction of variable renewable energy sources like solar panels~\cite{BRIGHT2018118satellitederived,Khalyasmaa2019PredictionOS} and wind turbines~\cite{Alencar2017DifferentMFwindpowergenerationcasestudy,LI2021121075powerpredictionwindturbinesgaussian} is an active field of research. %
Such models are usually based on weather models for mid- and long-term forecasts as well as, in case of solar, satellite data for short-term forecasts, that enable the observation and movement of clouds~\cite{KALLIOMYERS202068irradianceforecast}. 
Very short-term models with one-minute resolution can even be based on live video data using sky cameras\footnote{\url{https://solcast.com/utility-scale/solar-data-api/super-rapid-solar-farm-forecasts}}.
As wind and solar power production are known for their high variability, probabilistic prediction methods are especially common in this domain~\cite{VERBOIS2018313probabilisticforecastingsolarboosting,8982039probabilisticforecastingwindmixture}.

\subsection{Deriving the \emph{freep} capacity forecast}

Based on the previously generated forecasts, we now determine the main input to Cucumber's admission control: the \emph{freep} (\underline{f}ree \underline{REE}-\underline{p}owered) capacity forecast.

For this, we first calculate the REE forecast $P_\text{ree}$.
If no probabilistic forecasting was used to generate the power production $P_\text{prod}$ and consumption $P_\text{cons}$ forecasts, we can directly define $P_\text{ree} = \max(0,\ P_\text{prod} - P_\text{cons})$.
If probabilistic forecasting was applied, we now have the possibility to decide that $P_\text{ree}$ should describe a more optimistic or more conservative view of the future and hence manipulate the behavior of the admission control policy.

However, we need to differentiate between two kinds of probabilistic forecasts.
The first contains actual probability distributions for each forecasted observation, which, in practice, is mostly implemented as ensembles of non-deterministic single-value predictions.
In this case, the simplest way to build a joint distribution $P_\text{ree}$ is by randomly sampling from both distributions and subtracting the returned values for power production and consumption.
We can then use the quantile function $Q$
to determine a concrete single-valued time series.
\begin{equation}
    P_\text{ree}^{\,\alpha} = \max(0,\ Q(\alpha,\,P_\text{ree}))
\end{equation}
where $\alpha \in [0,1]$ determines how \emph{optimistic} (big $\alpha$) or \emph{conservative} (small $\alpha$) our forecasts are. 
For example, $P_\text{ree}^{\,0.95}$ returns the 95\textsuperscript{th} percentile of $P_\text{ree}$. 

In the second case, one or both forecasts do not contain the actual distributions but only values for a number of pre-initialized quantiles, usually the median and an upper and lower bound like the 10\textsuperscript{th} and 90\textsuperscript{th} percentile.
In this case, we propose a fall-back method as we cannot simply join the distributions:
\begin{equation}
    P_\text{ree}^{\,\alpha'} = \max(0,\ Q(\alpha,\,P_\text{prod}) - Q(1-\alpha,\,P_\text{cons}))
\end{equation}
where $\alpha'$ can only take certain values determined by the pre-initialized quantiles. Note that using this equation $\alpha'$ holds the same semantic value as $\alpha$ (e.g. big $\alpha'$ represents optimistic forecasts) but no guarantees of actual probability.
In the following, we use $\alpha$ and $\alpha'$ interchangeably.

Using the forecasts for computational load $U_\text{pred}$ and availbale REE $P_\text{ree}^{\,\alpha}$ we can now compute the \emph{freep} capacity forecast $U_\emph{freep}$, which determines how much of the free capacity in the future can be powered using only REE:
\begin{equation}
U_\emph{freep} = \min(
\overbrace{\rule{0pt}{3ex}{1 - U_\text{pred}}}^{\textstyle U_\text{free}},\
\overbrace{\rule{0pt}{3ex}{\frac{P_\text{ree}^{\,\alpha} - P_\text{static}}{P_\text{max} - P_\text{static}}}}^{\textstyle U_\text{reep}}
)
\end{equation}\label{eq:uflex}

The \emph{freep} capacity forecast is defined as the minimum of $U_\text{free}$, the expected free capacity of the node, and $U_\text{reep}$, the expected fraction of capacity that could be REE-powered.
If $U_\text{pred}$ is a probabilistic forecast, it first has to be converted to a single-valued time series, for example using $Q(0.5, U_\text{pred})$.
The equation for $U_\text{reep}$ depends on the used power model and it was derived by rearranging the linear power model from~\autoref{eq:power}.

\subsection{Admission Control Policy}

\begin{wrapfigure}{r}{0.52\textwidth}
  \vspace{-1.3cm}
  \begin{center}
    \includegraphics[width=0.50\textwidth]{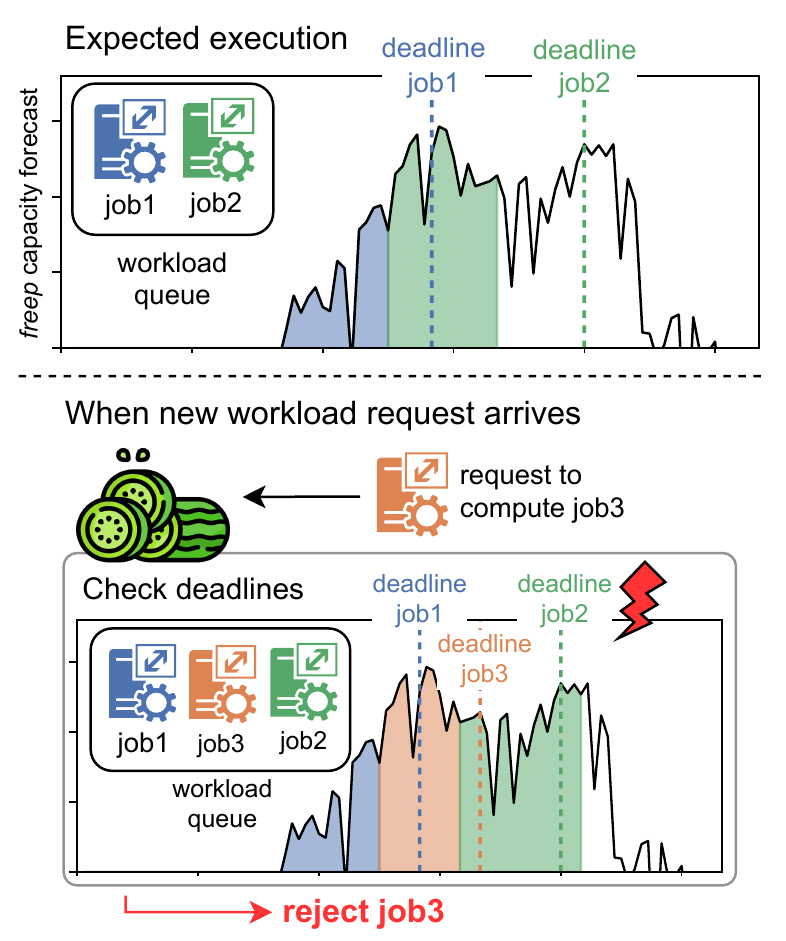}
  \end{center}
  \vspace{-0.5cm}
  \caption{Cucumber rejects workloads if it expects any future deadline violations using the \emph{freep} capacity forecast.}
  \vspace{-0.5cm}
\end{wrapfigure}

Cucumber admits workload requests based on the above derived \emph{freep} capacity forecast and the amount, size, and deadlines of already queued workloads.
For this, all workload requests are expected to provide a job size estimate and a deadline.
In practice, deadlines are often provided directly by users or services or can be derived from, for example, application graphs.
Estimating the size of jobs is a common problem in scheduling and is usually performed based on previous executions of the same or similar workloads.
In the current approach, we do not consider uncertainty in job size estimates, parallelism, or additional resource constraints besides computational load, like memory.
However, Cucumber can be extended to consider such factors.

The approach is agnostic to the applied scheduling mechanism, including multiple levels of priority or preemptive workloads, as long as it can be reliably modeled with the available information.
For every incoming request, Cucumber models the expected processing of the queue if the workload was accepted and evaluates if any deadlines are being violated.
That is, for each queued workload it progresses in time on the \emph{freep} capacity forecast until the expected (remaining) workload size is covered and then checks if the workload's deadline was violated.
If any violation occurs, the request gets rejected, otherwise accepted.

Depending on the number of workload requests and the average queue length, this basic algorithm can become computationally inefficient, since a re-evaluation has to take place for each request.
However, performance issues can be mitigated in many ways, for example by grouping jobs with the same or similar deadlines and only evaluation violations per group.
Moreover, different heuristics can be applied to decrease the number of re-evaluations, like caching the remaining time and capacity of each group until their deadline, and only performing a full re-evaluation once violations become likely.
Concrete performance adjustments depend on the nature of the underlying system, such as the level of parallelization as well as frequency, distribution, and kind of incoming workloads.

\subsection{Limiting Power Consumption at Runtime}

To ensure that accepted workloads run on REE only, their resource usage needs to be limited at runtime.
In practice, there are several ways to approach this, including adjustments of hardware power and speed settings like dynamic voltage and frequency scaling (DVFS).
Nevertheless, to propose a simple approach, modern high-level tools or resource orchestration solutions allow for conveniently controlling the usage of resources such as CPU or GPU.
For instance, the CPU usage of a process can be limited using tools like \emph{cpulimit}.
Likewise, frameworks like Docker and Kubernetes have built-in flags for limiting CPU usage by adapting the settings of a container’s \emph{cgroup}.
As load $U$ and available REE $P_\text{ree}$ can be measured periodically at runtime to derive the current $U_\text{gec}$, such tools can be used to adjust the node's power consumption to the correct level without inferring with the time-critical baseload.
However, the suitability of this simple approach depends highly on the concrete environment and more sophisticated measures might be needed in certain scenarios.

Even when performing admission control at a low $\alpha$ (meaning in conservative mode), conditions at runtime might still be worse than expected.
If less REE is available than forecasted, the previously described power limiting could lead to deadline violations of accepted jobs, although there is free computational capacity available.
While this behavior might be acceptable in some environments, usually it is more important to meet promised deadlines than ensuring that no grid energy is used at all.
To mitigate violations, Cucumber uses the \emph{freep} capacity forecasts at runtime to periodically evaluate whether the currently active jobs can still meet their deadlines.
If a running job is expected to violate its deadline, we temporarily stop power limiting and finish it using all free capacity $U_\text{free}$.
Since also load forecasts are uncertain, deadline violations still cannot be completely ruled out, but will be mitigated as effectively as possible based on the current state of knowledge.

\section{Evaluation}\label{sec:evaluation}

We evaluate Cucumber on real datasets over the course of two weeks (January 18-31) using the discrete-event simulation framework SimPy.
In total, 36 experiments were conducted: Six admission control policies (three baselines and Cucumber at $\alpha \in \{0.1, 0.5, 0.9\}$) in two scenarios at three solar sites each.
All data and simulation code are publicly available as mentioned in \autoref{sec:intro}.

\subsection{Experimental Setup}

We want to upfront explain some simplifications we made in our simulation-based evaluation.
First, we assume that the reported size of workload requests is always correct, while in practice runtime estimates are often noisy.
Yet, we consider this a problem not addressed by Cucumber.
Second, we do not explicitly model parallelism but process the workload queue next to the time-critical baseload in sequential order using non-preemptive earliest deadline first (EDF) scheduling.
Third, we do not model the energy demand of Cucumber itself.
However, we expect its overhead to be very small as forecasts are only updated every 10 minutes and the admission control itself can be implemented efficiently.

\subsubsection{Admission Control Policies}

We evaluate six admission control policies for each of the below-described scenarios and solar sites.
If deadlines are violated, jobs are not canceled but continue to run until they are completed.

\begin{itemize}
	\item \emph{Optimal w/o REE} accepts workloads using perfect forecasts for $U_{pred}$ but without considering the availability of REE. It declares the upper bound for accepted jobs without deadline misses but accepts high grid power usage.
	\item \emph{Optimal REE-Aware} accepts workloads using perfect load and renewable energy production forecasts. It declares the upper bound for accepted jobs without deadline misses and without any grid power usage.
	\item \emph{Naive} accepts workloads only if there is currently REE available and there is no other workload in process. This approach does not rely on forecasts.
	\item \emph{Conservative}, \emph{Expected}, and \emph{Optimistic} describe Cucumber admission control using realistic forecasts at $\alpha \in \{0.1, 0.5, 0.9\}$, respectively.
\end{itemize}

\subsubsection{Scenarios}

\begin{figure}[h]
    \centering
    \includegraphics[width=0.6\textwidth]{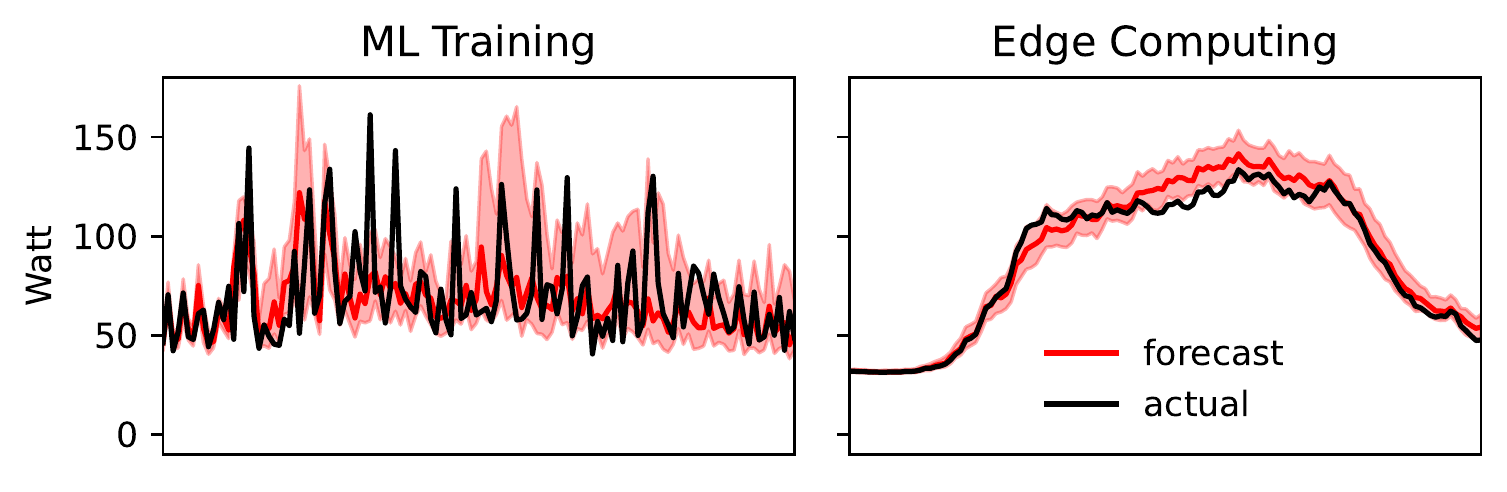}
    \includegraphics[width=0.9\textwidth]{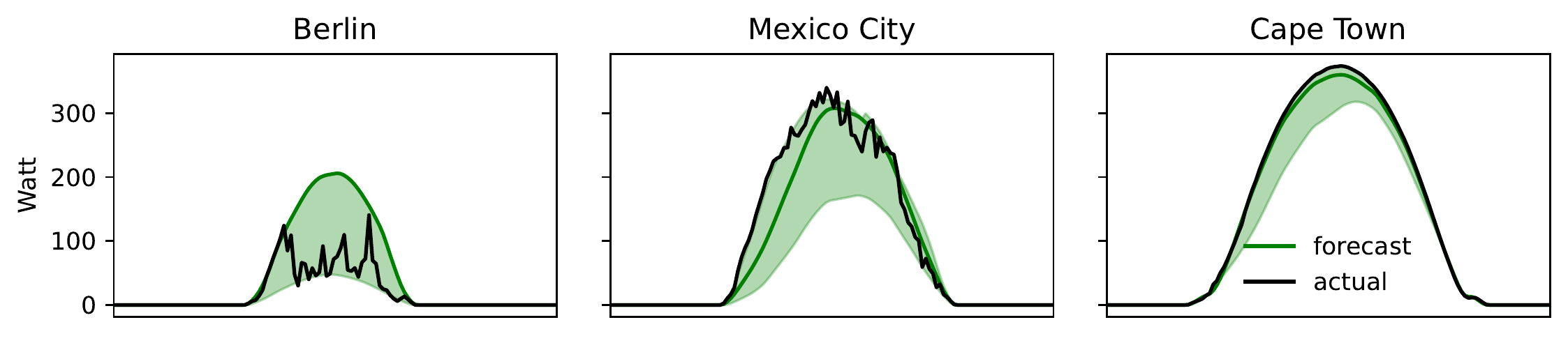}
    \caption{In red: actual and forecasted baseload power consumption in both scenarios at an exemplary day. In green: exemplary power production at the three solar sites.}
    \label{fig:datasets}
\end{figure}

We define two scenarios where each consists of a high-priority baseload and a number of workload requests.
Exemplary baseload patterns are depicted in \autoref{fig:datasets}.
Since, to the best of our knowledge, trace datasets with information on the delay-tolerance of workloads do not exist yet, we modeled both scenarios based on related real-world datasets:

\begin{enumerate}
    \item \emph{ML Training} is based on the \emph{cluster-trace-gpu-v2020} dataset from the Alibaba Cluster Trace Program\footnote{\url{https://github.com/alibaba/clusterdata}}, which contains two months of traces from a GPU production cluster~\cite{Alibaba_Data_2022}. 
   	Baseload is modeled using tasks labeled as \emph{worker}, which are highly variable and hard to predict.
   	Each of the 5477 delay-tolerant workload requests corresponds to an \emph{xComputeWorker} task in the dataset.
    The size of workloads is determined by the \emph{plan\_gpu} property and each workload has to be finished by midnight the day it was issued, meaning deadlines can be anywhere from 0 to 24 hours.
    \item \emph{Edge Computing}: is based on the NYC Taxi Trip dataset\footnote{\url{https://www1.nyc.gov/site/tlc/about/tlc-trip-record-data.page}} from Dec 2020 and Jan 2021. Baseload is modeled on the number of yellow taxi rides, which is highly seasonal. The 2967 workload requests correspond to long-distance green taxi rides: Every green taxi ride over 10\,km length emits a job at \emph{lpep\_pickup\_datetime} which has to be computed until \emph{lpep\_dropoff\_datetime}. The median deadline is 41 minutes. All jobs have the same size.
\end{enumerate}

We generated baseload forecasts by training a DeepAR~\cite{salinas2020deepar} probabilistic forecasting model\footnote{DeepAR parameters: GRU, 3 Layers, 64 nodes, 0.1 Dropout; 20-30 minutes training time on commodity hardware} on the first 1.5 months of data to then generate 24-hour forecasts with a 10-minute resolution for every 10-minute step in the last two weeks of the datasets. 
Note, that the arrival rate of workload requests is not forecasted by Cucumber.
Power consumption forecasts are derived using \autoref{eq:power} with $P_\text{max} = 180\,W$ and $P_\text{static} = 30\,W$.

\vspace{-0.7cm}

\subsubsection{Solar Sites}

We assume every compute node has access to a solar panel with 400\,W peak production.
We collected real solar power production forecasts using the Solcast\footnote{\url{https://solcast.com}} utility-scale API during the second half of January 2022.
Like load forecasts, the solar forecasts cover 24 hours in 10-minute resolution each and were generated in 10-minute intervals.
Each forecast contains the median as well as the 10\textsuperscript{th} and 90\textsuperscript{th} percentile of expected energy production for each time step. 
To evaluate the effectiveness of our approach at different geographical locations and during different seasons, we gathered forecasts at three different sites located at different continents and latitudes:

\begin{enumerate}
	\item \emph{Berlin} during winter (8 hours of daylight; 2 hours of sunshine)
	\item \emph{Mexico City} during the dry season (11 hours of daylight; 7 hours of sunshine)
	\item \emph{Cape Town} during summer (14 hours of daylight; 11 hours of sunshine)
\end{enumerate}	

For orientation, the roughly expected hours of daylight and sunshine in January at each site are listed in parentheses.
Exemplary values for each site are displayed in \autoref{fig:datasets}.

\subsection{Results}

\begin{figure}
    \centering
    \includegraphics[width=\textwidth]{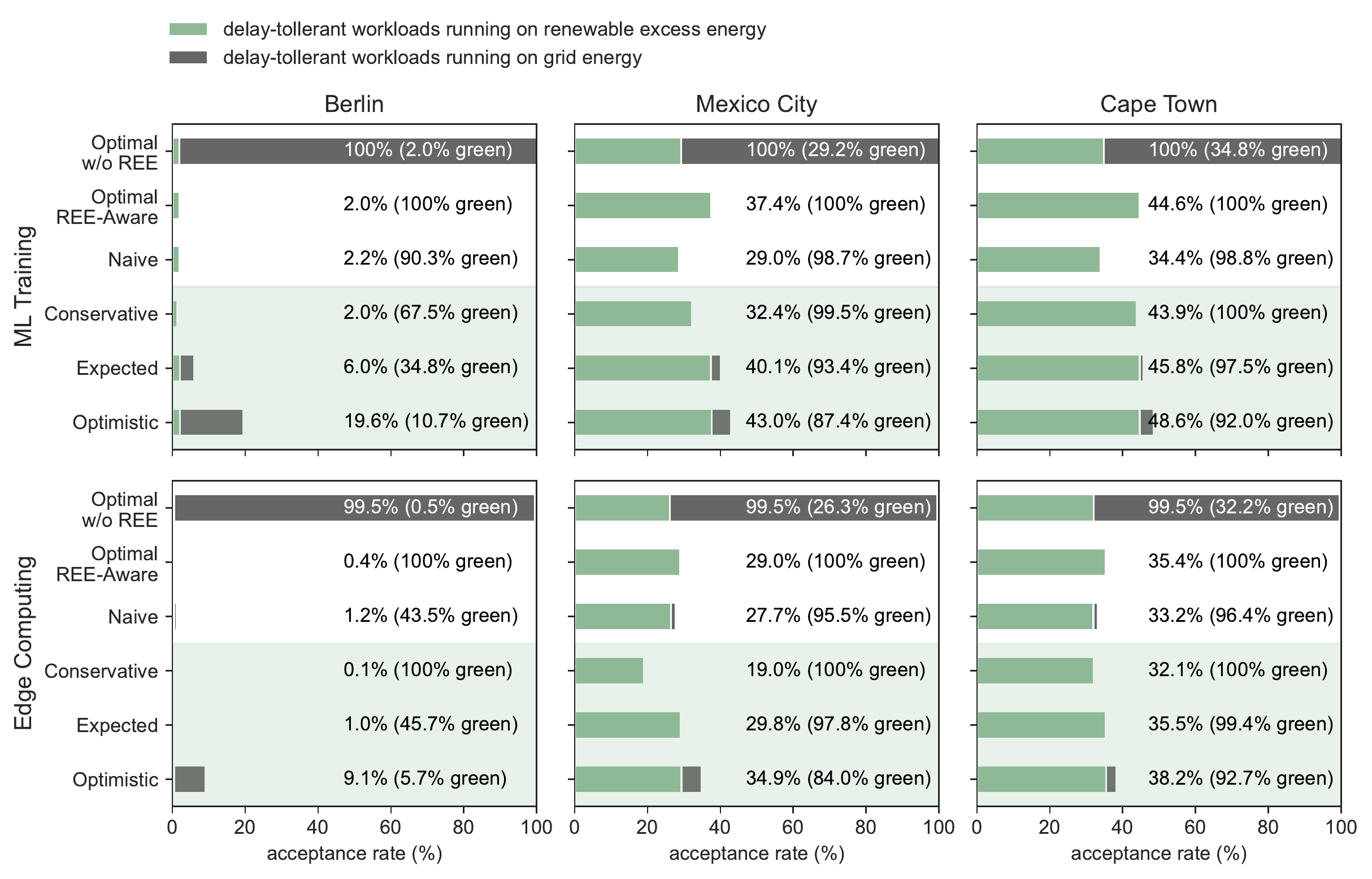}
    \caption{Acceptance rate of workload requests and the fraction of these workloads that was actually powered via REE during execution (green).}
    \label{fig:evaluation}
\end{figure}

For each experiment, we report the admission control acceptance rate and the fraction of REE that was used to actually power the workloads.
\autoref{fig:evaluation} illustrates the results.

\vspace{-0.3cm}

\vspace{-0.3cm}

As expected, \emph{Optimal w/o REE} accepts almost all workload requests at the cost of requiring a substantial amount of grid energy. 
Worth mentioning is the constant acceptance rate of 100\,\% across all experiments of the ML Training scenario, which is a result of the rather relaxed deadlines.
The stricter deadlines in the Edge Computing scenario lead to a slight decrease in acceptance rates.
Both baselines utilize perfect forecasts but only \emph{Optimal REE-Aware} considers available REE, which is why is does not use any grid energy across all experiments.

We observe that there was barely any REE available at the Berlin solar site during the observed period.
Even \emph{Optimal REE-Aware} accepts only a maximum of 2\,\% of all workloads.
Since the uncertainty and error rate of solar forecasts is extremely high at the Berlin site, only \emph{Conservative} forecasts achieved comparably low grid power usage.
Admission control based on \emph{Optimistic} and \emph{Expected} forecasts resulted in very low REE usage of 5.7\,-\,10.7\,\% and 34.8\,-\,45.7\,\%, respectively.
Under such conditions, the usage of a forecast-based admission control policy such as Cucumber can hardly be justified, as it does not show improved performance compared to a \emph{Naive} approach.

However, in Mexico City and Cape Town, which had a lot longer days and better weather during January than Berlin, Cucumber clearly outperforms the \emph{Naive} admission control, which achieves 31.1\,\% acceptance rate at 97.3\,\% REE usage in average.
Cucumber's \emph{Expected} case configuration maintains almost the same REE usage (97.0\,\%) but increases the acceptance rate to 37.8\,\%, while the \emph{Conservative} configuration manages 99.9\,\% REE usage at an acceptance rate of 31.9\,\%.
The trade-off when tuning the forecasts is clearly visible: 
While \emph{Conservative} admission control results in almost perfect REE coverage, the acceptance rate was on average 18.5\,\% lower.

\begin{figure}
    \centering
    \includegraphics[width=0.95\textwidth]{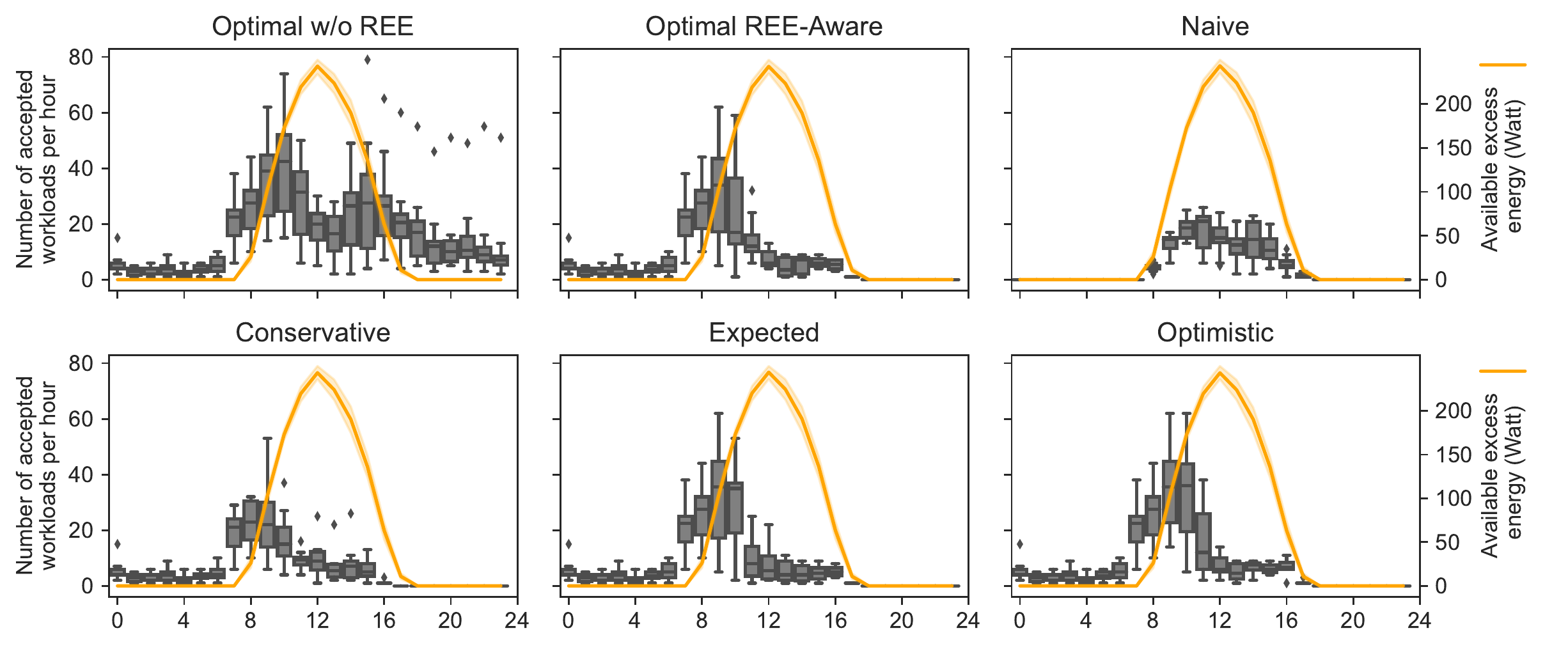}
    \caption{Aggregated number of accepted workloads per hour for all admission control policies during the ML Training scenario in Mexico City. The orange line indicates the average solar production during a day.}
    \label{fig:boxplots}
\end{figure}

\autoref{fig:boxplots} depicts the aggregated number of jobs per hour for an exemplary solar site on the ML Training scenario (all deadlines are midnight).
We observe that the acceptance rate over time differs strongly between the different approaches: 
Considering that \emph{Optimal w/o REE} describes all workloads that can be accepted without deadline violations, \emph{Optimal REE-Aware} describes the optimal subset that can be computed using only REE.
The \emph{Naive} approach cannot exploit this potential, as it only accepts workloads once there is REE available.

The Cucumber admission control, on the other hand, is based on forecasts of REE and hence already accepts workloads before the sun is rising.
It can be observed, that in the the \emph{Expected} case's behaviour is close to the optimal case and almost all jobs before 11\:am get accepted. 
After that, the number of accepted jobs per hour falls drastically since the forecasted solar energy until midnight is already reserved by queued workloads and forecasts in Mexico City are comparably precise.
In \emph{Conservative} mode, Cucumber is more cautious and accepts fewer jobs during early morning hours.
However, it accepts additional jobs throughout the day as uncertainty decreases when progressing in time.

We note that \emph{Optimistic} forecasts barely increase REE usage compared to Expected forecasts in most experiments.
For example, the acceptance rate for the Edge Computing scenario in Mexico City went up by 16.3\,\%, but the REE usage by only 0.5\,\%, meaning that almost all additionally accepted jobs were powered by grid energy.
Furthermore, we note that the \emph{Optimistic} experiments resulted in 1, 5, and 7 deadline misses in the Edge Computing scenario (which has tight deadlines), while none of the other configurations caused any deadline misses.
We conclude that users should pick $\alpha > 0.5$ with caution.

\section{Conclusion}\label{sec:conclusion}

This paper presents Cucumber, a configurable admission control policy for re\-source-con\-strained compute nodes with on-site renewable energy sources.
Cucumber accepts delay-tolerant workloads to increase REE utilization through probabilistic multistep-ahead forecasts of computational load, energy consumption, and energy production.
Our simulation-based evaluation uses real solar production forecasts for Berlin, Mexico City, and Cape Town and compares different configurations of our approach with baseline policies on two exemplary scenarios.
The results show, that Cucumber's default configuration shows similar acceptance rates than the optimal case baseline while achieving an REE coverage of 97.0\,\% on average in Mexico City and Cape Town.
Conservative admission results in almost perfect REE coverage at a 18.5\,\% reduced acceptance rate.

For future work, we plan to implement Cucumber in a hardware testbed to study its behavior and computational overhead under realistic conditions.
Furthermore, we want to extend the approach to also consider available energy story and make Cucumber part of a decentralized architecture that exploits the spatio-temporal availability of REE in a distributed system via local decisions.

\section*{Acknowledgments and Data Availability Statement}

The datasets and code generated and analyzed in this paper are available in the Figshare repository: \urlstyle{same}\url{https://doi.org/10.6084/m9.figshare.19984556}~\cite{figshare_artifact}.

We sincerely thank Solcast for the uncomplicated and free access to their solar forecast APIs.
This research was supported by the German Academic Exchange Service (DAAD) as ide3a and the German Ministry for~Education and Research (BMBF) as \mbox{BIFOLD} (grant 01IS18025A) and Software Campus (01IS17050).

\bibliographystyle{splncs04}
\bibliography{bibliography}

\end{document}